# The physical foundation beneath protein generative modeling

Tianyu Lu[a], Po-Ssu Huang[a]

[a]Department of Bioengineering, Stanford University, Stanford, California, USA.
Corresponding Author: Huang, Po-Ssu

## Highlights

- Generative models map their latent representations to physically plausible regions of design space

- Desired inference-time distribution is shifted from training distribution but the underlying physics remains constant

- Shaping the model's physical world view can be done with energy-based losses, architectural regularization, and experimental data feedback

## Abstract

Physical equations in protein modeling appear to have been replaced by generative models trained directly on structure data. By learning a mapping from noise to data, sampling *de novo* protein structures has become much more efficient. However, such models can also learn non-physical features and break down with out-of-distribution settings which are typical in protein design campaigns. In this review, we highlight where the physics persists in protein generative modeling pipelines and the issues that linger when physically relevant components of a macromolecular system are left unmodeled. We present a perspective that respecting the underlying physics of macromolecular systems, increasingly through learned representations that are physically grounded and updating generative models with experimental data, is foundational to generative modeling for functional protein design.

## 1. Introduction

The modeling of macromolecules has historically required physical equations: force fields, energy functions, rotamer libraries, and molecular-mechanics terms [1, 2, 3]. The Rosetta suite of tools, at its core, is the Rosetta energy function: van der Waals packing, hydrogen bonding, electrostatics, implicit solvation, and data-driven terms from the Protein Data Bank (PDB) [3]. The terms of the energy function were partially tuned to distinguish decoys, i.e. synthetic structures, from crystallographic structures. Recently, however, the field has shifted away from physical equations towards generative models trained on structural data [4, 5, 6, 7]. One distinguishing advantage is their ability to sample from the distribution of structures much more efficiently than the steps required using physics-derived methods. Another more critical feature is the ability to implicitly capture relations in the data that resemble physical interactions by learning. Neural network weights can acquire information correlated with physical equations when fitted to the data distribution. In presenting macromolecular data to the network, the information can be grounded in physical inputs such as coordinates and bond geometries, but can also be non-physical inputs which could lead the model to learn non-physical correlations. In this review, we discuss the issues that arise when physics is not modeled and highlight the current generative modeling paradigm where physics-based modeling remains important.

# 2. Generative models as physical priors

Generative models learn a distribution of its training data. By training on experimentally determined protein structures, the physics which governs physically plausible structures are implicitly modeled. The probability mass defined in the model is concentrated where folded proteins have been observed, and a well-trained model produces samples drawn from that region without evaluating any explicit energy. The distribution it approximates is the empirical distribution of deposited structures, which are dominated by proteins that fold, crystallize, and are often cryo-cooled, which is related but not equivalent to the distribution of structures in solution or in their native cellular environments at relevant temperatures, pH, and solvents. Such models serve as priors for protein design tasks which seek to draw samples from a bespoke region of the total space of physically plausible structures [8]. Many design tasks necessitate the proposal of a novel sequence or structure to achieve an objective previously not observed in the training data. The proposed sample therefore resides, to some degree, outside the training data distribution, and the success of a design campaign depends on the generalizability of the models used [9].

# 3. Steering the prior towards a design objective

## 3.1. Physical terms as filters and guidance

In design, the prior distribution learned by a generative model is almost always different from the distribution from which we wish to sample. For example, the distribution of all protein-protein interfaces is different from the distribution of high affinity interfaces with minimal off-target binding. This distribution shift can be approximated by *post hoc* filters. BindCraft applies Rosetta-based physical terms [10] to filter generated designs to ground them in physical plausibility, together with confidence metrics from black-box structure prediction based oracles [11], to enrich samples that are likely to be high affinity binders. However, structure prediction models are often only trained on positive data. In the binder design setting, confidence metrics capture the confidence of the predicted pose, implicitly conditional on the assumption that the interaction is real. *Post hoc* filters are bounded by what the generator already proposes and inherits the generalizability limits of the oracles. A meta-analysis of 3,766 experimentally characterized binders show that physics-based metrics provide additional predictive power to sort *in silico* binders from non-binders compared to structure prediction metrics alone, such as ipSAE [12]. Physical terms can also be applied during the generation process of the sample by steering the denoising process with Feynman-Kac resampling, beam search, or reinforcement learning against non-differentiable physical validity rewards [7, 13, 14]. Physical plausibility from the perspective of a predictive model could also be transferred to generative models by finetuning predictive models as generative models, such as RFdiffusion which builds on RoseTTAFold [15, 16]. Whether using physical terms as filters or guidance, their intended effect is to shift the prior learned by the generative model towards a more bespoke distribution for a given design objective (Figure 1).

## Limitations of structure prediction models for ranking designs

Using folding models as oracles may have unintended consequences. For example, charge inversion mutations, which were expected to electrostatically abolish an interaction with the negatively charged ATP, did not perturb the predicted ligand binding pose nor substantially reduce the confidence [17]. These adversarial examples suggest that co-folding models tend to interpolate within its training data distribution and do not adequately capture physical expectation [18, 19]. Additionally, use of prediction-derived metrics such as ipSAE often results in ranking false positive designs highly, according to the general distribution of interfaces, while failing to distinguish subtle incompatibilities within the samples (Figure 2). These issues highlight the so-called activity-cliff or rugged landscape of proteins, where small changes to the sequence or structure lead to dramatic changes in its physical plausibility or function. A strategy to circumvent this is to introduce negatives during the inference time of generative models. In autoguidance, a generative model is guided with a “bad” version of itself, typically an under-trained model which produces samples of poor quality, such that sampling is pushed away from the failure modes of the under-trained model [20]. This approach is similar in motivation to tuning the Rosetta energy function to separate decoys from natives, but has the advantage of being more flexible and is not subject to the expressibility constraints imposed by physical equations with much fewer parameters. A similar behavior occurs on the generalization of both design models and predictive models on longer length proteins. While subject to the same physics, longer proteins tend to underperform in designability metrics [8, 21]. The struggle with length generalization points to the same phenomenon, that physics is not adequately captured by generative models or by structure prediction oracles, since length is a statistical property of the training set rather than a physical variable.

# Grounding models in physics-based representations and architectures

## Using physical energies to organize an all-atom latent space

A recent class of models attempts to focus the network on relevant physical information by redefining the representations of molecules to more fundamental building blocks, typically a collection of atoms within a few Ångstroms of range [22, 23]. Efforts in machine learning interaction potential (MLIP) development have directed their learning systems towards predicting inter-atomic energy by fitting models to pre-computed classical or quantum calculations [24, 25], with the goal of speeding up traditionally slow and expensive calculations. The emphasis on local interactions is also critical for model generalizability. Models that only use structure-based training objectives can readily complete structural reconstruction tasks by memorizing non-generalizable features of a protein. In contrast, leveraging information from physically-grounded features is likely to be more generalizable to unseen problems. However, there is an informational gap between the localized atomic features and the global fold of a protein. A model that directly bridges this gap could overcome the performance cliffs observed due to memorization of non-physical features by assessing its ability to use strictly local features to reconstruct the global fold. Recent work showing the convergence of internal representations of scientific foundation models hint at a common physical structure learned by otherwise dissimilar models [26, 27]. To disentangle the bottom-up physical features and the context associated with the global structure, SLAE [23] leverages an asymmetrical encoder-decoder architecture. It attempts to project global structure to a latent space by using only localized information to encode it, while simultaneously trying to achieve accurate reconstruction with a transformer-based decoder. Ablation analyses showed that predicting energy by the decoder was essential during model training for the localized all-atom representation to perform well simultaneously at global fold reconstruction and at downstream tasks such as NMR chemical shift prediction, function prediction, and all-atom sidechain reconstruction.  Remarkably, the model can achieve physically accurate and smooth linear interpolation between protein states. This hints at a well-organized latent space: local all-atom representations now need to be organized along physically meaningful axes by virtue of the asymmetric decoder's energy supervision task, jointly with global structure reconstruction. This prevents memorization of spurious features of the training data. This approach can be viewed as a hybrid between purely data-driven representations and closed-form energy functions, inheriting both the generalizability from the physical supervision and the expressivity of representation learning. A similar approach is the affinity module in Boltz-2 where the organization and information content of the denoising module representations are updated during training to better predict binding likelihoods and affinities [28].

## Physics-inspired architectural regularization

Physical models also inspire architectures of protein design models. The Potts model, originally developed for modeling interacting spins on a crystalline lattice, has had successful applications in protein design [29]. Earlier methods rely on co-evolutionary information derived from multiple sequence alignments to infer the Potts model parameters. Recently, models such as TERMinator, ChromaDesign, and Caliby parameterize the Potts model with outputs of a neural network [6, 30, 31]. The decision to use a Potts model to parameterize the distribution of sequences from which to sample may seem counterintuitive given the flexibility of autoregressive models, as the latter can capture any arbitrary distribution over sequences, including higher-order correlations beyond second order. However, several works indicate that second order terms are necessary and sufficient to capture protein structure and function [32, 33, 34]. Thus restricting to only first and second order terms in a Potts model can be viewed as a form of physics-inspired architectural regularization. It serves as a parameterization bottleneck to prevent the modeling of possibly spurious higher order terms, which may explain the drop in designability for ProteinMPNN compared to ChromaDesign and Caliby, especially for longer protein lengths [31].

# Connecting generative model hypotheses with experiment

## Distilling the necessary and sufficient physical conditions for functional protein design

Achieving successful “zero-shot” design highlights the successful distillation of physical principles into necessary and sufficient conditioning inputs to modern protein design tools. A *de novo* designed protein was able to bind the drug exatecan which has no prior structure in the PDB nor the Cambridge Structural Database (CSD) by iterating between protein-ligand co-folding with AlphaFold3 and sequence redesign with LASErMPNN, starting from ligand interactions derived from COMBS [35]. A multi-objective genetic algorithm was used to design a *de novo* zinc-dependent phosphatase [36] by iterating between a metal site predictor Metal3D [37], sequence redesign with Caliby, and subsequent mutation and repacking with LigandMPNN [38], starting from a Protpardelle backbone [39]. In a *de novo* serine hydrolase design campaign, PLACER was used to model the preorganization of a DFT-derived active site with scaffolding done by RFdiffusion [15, 40, 41].

## Implications of unmodeled physical components

As with any modeling approach, there are components of the system that are not modeled, including explicit solvent, energy barriers due to desolvation of the binding pocket, and the entropic versus enthalpic components of binding, among other aspects. Other external constants are also not frequently modeled, including stability at some given pH and temperature, desired protonation states, and competition with off-targets. In many cases, the choice to not model a given component is deliberate, for example hydrogens only being resolved in very high resolution crystal structures. In other cases, the missing components are integral to the success of a design campaign. For example, in a serine hydrolase design campaign, the activated water in the second step of the double-displacement ester hydrolysis reaction is not explicitly modeled while being essential to the enzyme mechanism [40]. In binder design, the often desired selectivity criterion is often not explicitly modeled. With a motivation similar to autoguidance, AlloGen demonstrates that off-target states can be steered against during inference time [42]. Electrostatics form an integral part of physics-based design methods, where accounting for their long-range effect is a challenge for molecular dynamics simulations. However, the modeling of electrostatics is sparse in recent machine learning methods for design and standard MLIPs also do not account for long-range interactions [24]. The orientation of an electric field is thought to be the main contributor to rate acceleration in well-studied enzymes, where contributions to the electric field can come from distal sites of the protein [43, 44, 45]. Charge stabilization by an oxyanion hole is a consequence of this: in the serine hydrolase design campaign, mimicking natural ester hydrolases in constructing the oxyanion hole motif using the residue after the catalytic serine led to a five fold rate improvement [40]. When physical constants are not modeled, fallback to natural occurrence can be a viable strategy [46]. However, modeling electrostatics is dependent on accurate protonation states which exist as an equilibrium and are solvent and pH dependent, all of which are usually not modeled and so are added or predicted with tools decoupled from the generative model, such as applying molecular dynamics *post hoc* to validate rather than steering the generative process [47, 48]. This could explain why low temperature sampling, where models are altered during inference time to concentrate more probability density on its more confident regions, is almost always necessary to produce designable samples, as concentrating samples on the modes of the distribution effectively averages out the unmodeled latent variables in the model.

## Kinetics as an explicit design objective

Low temperature sampling exacerbates the tendency to sample overly stable designs while not explicitly modeling kinetics, where over-optimization for designability could ignore biologically relevant modes [8]. Most biologically relevant phenomena require a delicate balance between stability and instability [49], all while being constrained by physical plausibility. Broerman *et al*. considered kinetics as an explicit design objective by intentional design of clashes in a strained intermediate which are resolved by dissociation [50]. The dissociation rate can be increased by directing the force of the strained complex to shear along the axis of secondary structure elements, whereas clashes that require rotation to resolve showed slower dissociation rates. In this example, the prudent use of low temperature sampling with RFdiffusion was important to stabilize the fusion of the binder to the conformational switch, since that interaction must be more stable than the desired detaching interface, forcing the dissociation rather than rearrangement of the complex. Nonetheless, the process required manual specification of the strained states not accessible to generative models trained on ground states.

## Testing designs as plausibility queries in the physical world

A strategy to account for the difficult-to-model components is to include experimental data. This is the classic design-build-test-learn cycle in engineering, for example using experimental data to better fit Rosetta energy terms to reduce overpacking [51]. Generative models serve as priors which can be updated with experimental results using Bayesian optimization [52]. Each design is a hypothesis and queries physical reality, ideally to maximize information gain, typically by maximizing an upper confidence bound of an oracle. In this perspective, the goal of generative model development is to sufficiently map out a hypothesis space that is both easy to sample from and sufficiently expressive such that it is capable of responding to highly nonlinear experimental data. Beyond the baseline assessments of designability, it may be more instructive to evaluate model performance by how efficiently it can adapt its placement of probability density on high quality, but previously out-of-distribution examples. These rounds of iteration can also be entirely lab-driven, with a design serving as a progenitor in subsequent rounds of directed evolution [53], where all unmodeled latent physical constraints are jointly optimized by mutagenesis and selection.

### New assays to expand the modelable space

New developments in experimental methods can expand the currently observable protein properties amenable to modeling, often at higher throughput compared to previous methods. Using Tandem Mass Tag (TMT)- multiplexed bottom-up mass spectrometry, Martell *et al*. measured the extent of aggregation after exposure to acidic pH 4 or thermal stress at 50 °C and 75 °C of 18,987 protein domains, all concentration normalized to 10 mg/mL [54]. Ferrari *et al*. quantified residue level energy profiles of protein conformational fluctuations combined across pH 6 and pH 9 for 5,778 protein domains in parallel, distinguishing between all-or-nothing folded to disordered transitions from domains that have partially open intermediates, revealed by differences in hydrogen-deuterium exchange rates for protected or exposed backbone amide hydrogens [55]. Adapting the semi-automated protein production protocol from Qian *et al*. [56], Müntener and Abramson *et al*. scaled NMR to characterize the local fluctuations of 384 designed proteins, with spectrometer time being the bottleneck at 224 per week, with faster data acquisition rates possible at expression volumes scaled up to just 16 mL [57]. Passow *et al*. developed an amplicon/protein bead display method which can express and purify >100,000 proteins in one day, and demonstrated a 3-day protocol to quantify the expression levels and equilibrium dissociation constants of 18,000 protein variants [58]. Taken together, these new experimental techniques could enable incorporating protein energy landscapes, stability at a desired pH or temperature as explicit design objectives [59]. Coupling design with direct measurements of protein dynamics has the potential to directly train generative models on such data, building upon efforts to train on synthetic simulations of conformational ensembles [60].

## Conclusion

Macromolecules exist in a physical world. Despite the unique capabilities of learning systems in integrating information, modeling methods ultimately should respect the underlying physics. As more significant heights are being achieved in designing functional proteins, new solutions will lean more heavily on our ability to model the underlying physical principles. We outline here various strategies in practice today, mainly relying on model architecture and data curation. As lab-in-a-loop and various automation processes become more efficient and mature, data centric design tools will undergo dramatic transformation in the near future.

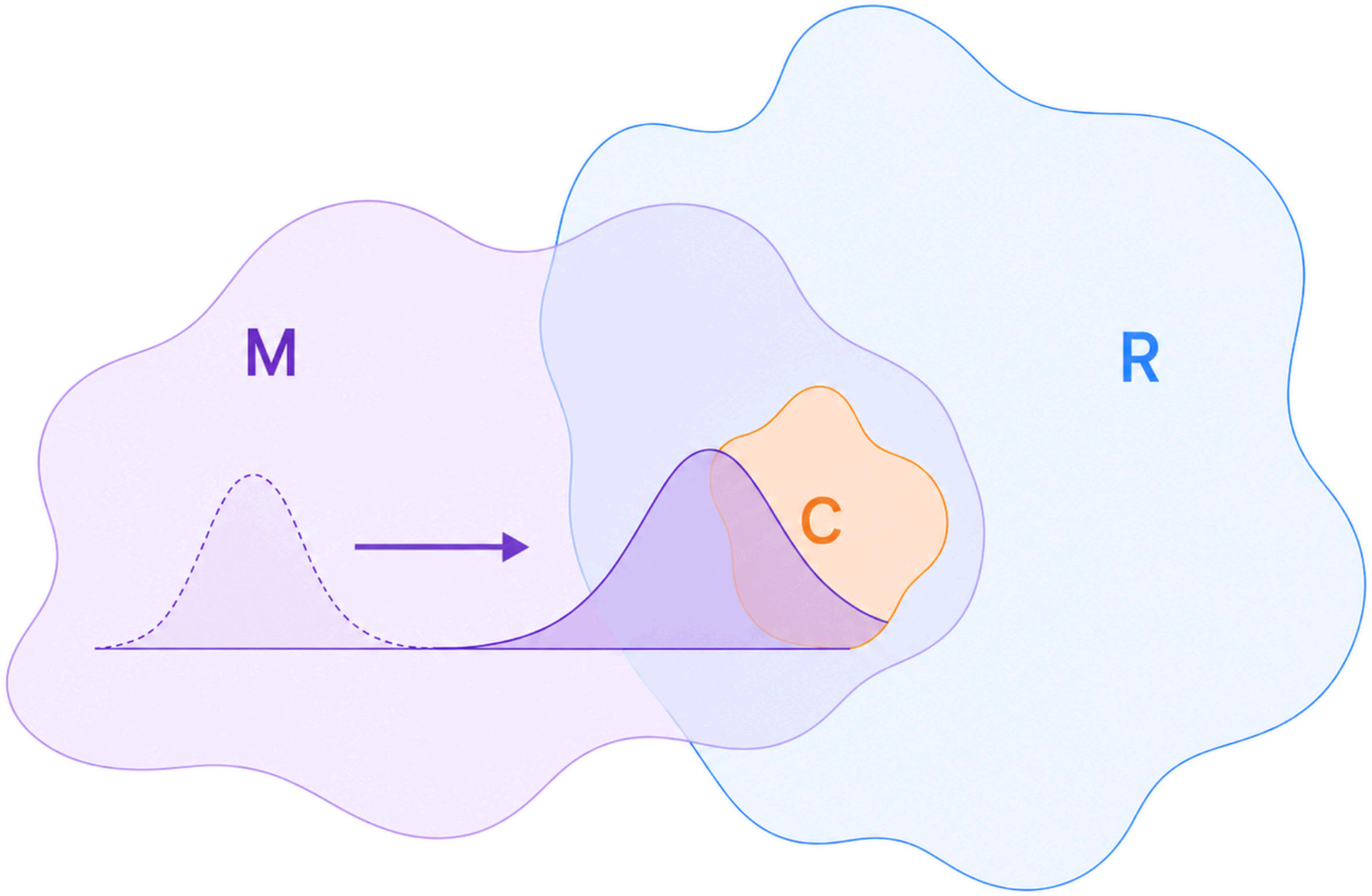

**Figure 1:** A depiction of the generative design process. **M:** Model distribution. **R:** physically plausible distribution. **C:** design objective distribution. The role of physical terms used in steering or as filters shifts the probability density to increase the proportion of samples that are physically plausible. Without proper specification of such conditions that are necessary and sufficient for an intended designed function, rejection sampling can be highly inefficient. With well-calibrated guidance or a correct set of conditions, the samples can be more enriched for the desired function.

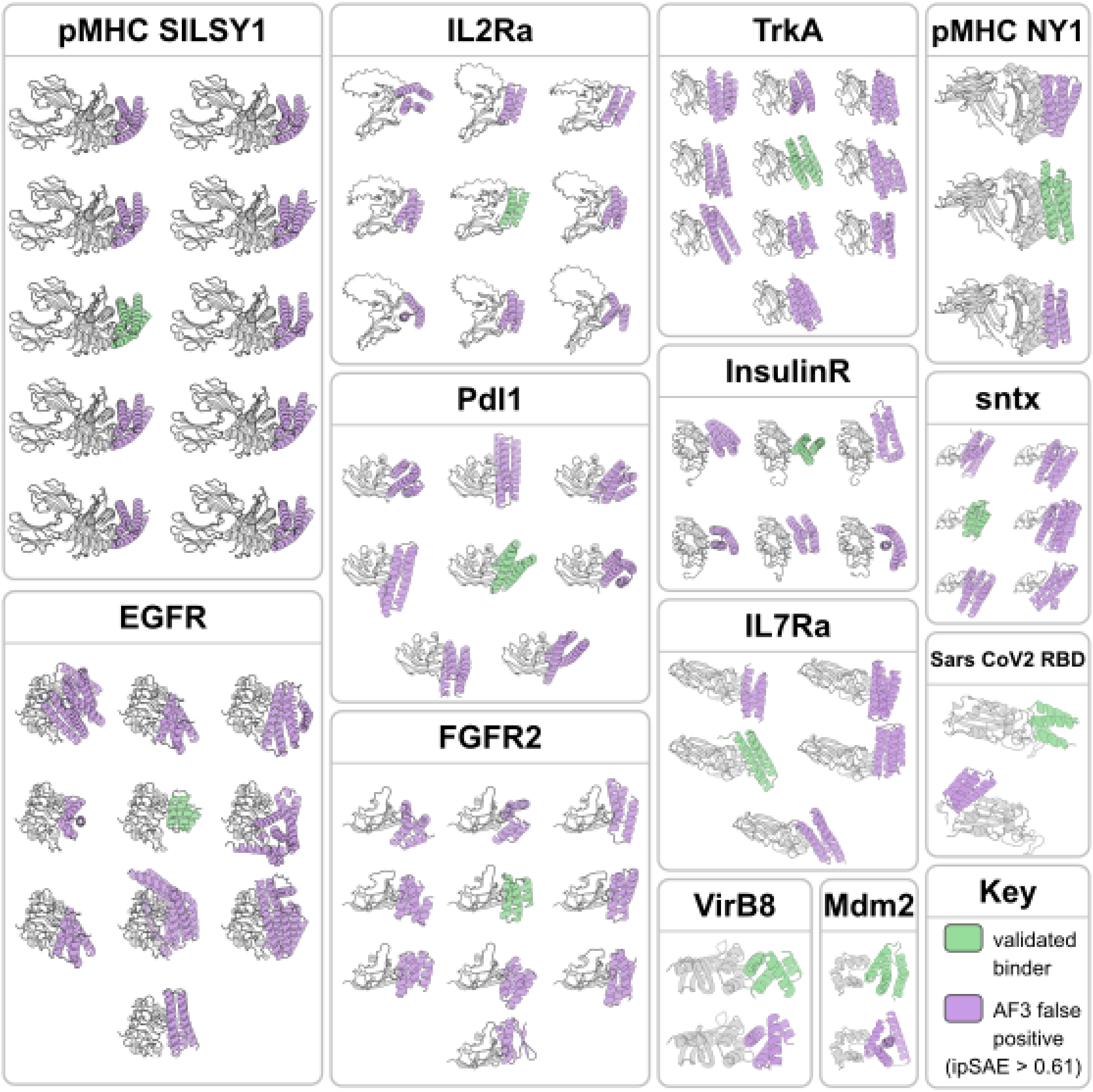

**Figure 2**: Structure prediction metrics are low precision. Using the AF3 [61] min ipSAE metric threshold of 0.61, a large proportion of designs predicted to be binders (> 0.61) are not binders when tested experimentally. The challenge is to distinguish between the true binders (green) to the target (grey) from the highly plausible designs which fail in experiment (purple). The number of false positives is proportional to the number of true positives given in Overath *et al.* [12].

## Acknowledgements

T.L. is supported by the Stanford Graduate Fellowship. This project is supported by NIH (R01GM147893 to P.-S.H.), Merck Research Laboratories (MRL) Scientific Engagement and Emerging Discovery Science (SEEDS) Program, and Stanford Medicine Catalyst. The views and conclusions contained in this document are those of the authors and should not be interpreted as representing the official policies, either expressed or implied, of the U.S. Government.

## Declaration of Interests

None.

During the preparation of this work, the authors used Claude Cowork for citation management. The authors reviewed and edited the output as needed and take full responsibility for the content of the published article.

## References and recommended reading

Papers of particular interest, published within the period of review, have been highlighted as:

* of special interest
** of outstanding interest

[1]. Poole AM, Ranganathan R. Knowledge-based potentials in protein design. Current opinion in structural biology. 2006 Aug 1;16(4):508-13.

[2]. Boas FE, Harbury PB. Potential energy functions for protein design. Current opinion in structural biology. 2007 Apr 1;17(2):199-204.

[3]. Leman JK, Weitzner BD, Lewis SM, Adolf-Bryfogle J, Alam N, Alford RF, Aprahamian M, Baker D, Barlow KA, Barth P, Basanta B. Macromolecular modeling and design in Rosetta: recent methods and frameworks. Nature methods. 2020 Jul;17(7):665-80.

[4]. Butcher J, Krishna R, Mitra R, Brent RI, Li Y, Corley N, Kim PT, Funk J, Mathis S, Salike S, Muraishi A. De novo design of all-atom biomolecular interactions with rfdiffusion3. bioRxiv. 2025 Nov 19.

[5]. Lin Y, Lee M, Vermani A, Jiang E, De Cooman S, Špeťko M, AlQuraishi M. Fast and Ultra-Capable Protein Design: Advancing the Frontier Through Atomistic SE (3)-Equivariance with Genie 3. bioRxiv. 2026 May 5:2026-05.

[6]. Ingraham JB, Baranov M, Costello Z, Barber KW, Wang W, Ismail A, Frappier V, Lord DM, Ng-Thow-Hing C, Van Vlack ER, Tie S. Illuminating protein space with a programmable generative model. Nature. 2023 Nov 30;623(7989):1070-8.

[7]. Didi K, Zhang Z, Zhou G, Reidenbach D, Cao Z, Cha S, et al.: Scaling atomistic protein binder design with generative pretraining and test-time compute. In Proc 14th ICLR 2026.

[8]. Lu T, Liu M, Chen Y, Kim J, Huang PS. Assessing generative model coverage of protein structures with SHAPES. Cell Systems. 2025 Aug 20;16(8).

[9]. Listgarten J, Jiang H. How artificial intelligence is reengineering protein engineering. Science. 2026 Apr 9;392(6794):159-66.

[10]. Pacesa M, Nickel L, Schellhaas C, Schmidt J, Pyatova E, Kissling L, Barendse P, Choudhury J, Kapoor S, Alcaraz-Serna A, Cho Y. One-shot design of functional protein binders with BindCraft. Nature. 2025 Oct 9;646(8084):483-92.

[11]. Jumper J, Evans R, Pritzel A, Green T, Figurnov M, Ronneberger O, Tunyasuvunakool K, Bates R, Žídek A, Potapenko A, Bridgland A. Highly accurate protein structure prediction with AlphaFold. nature. 2021 Aug 26;596(7873):583-9.

[12**]. Overath MD, Rygaard AS, Jacobsen CP, Brasas V, Morell O, Sormanni P, Jenkins TP. Predicting experimental success in de novo binder design: a meta-analysis of 3,766 experimentally characterised binders. BioRxiv. 2025 Aug 14:2025-08.

This meta-analysis of 3,766 experimentally characterized de novo binders shows that physics-based scoring adds predictive power for separating true binders from decoys beyond structure-prediction confidence metrics such as ipSAE alone. It provides quantitative evidence across different design campaigns that oracle confidence tend to be a low precision metric for binder design success.

[13]. Shen J, Zhou S, Che X. FKSFold: Improving AlphaFold3-Type Predictions of Molecular Glue–Induced Ternary Complexes with Feynman–Kac–Steered Diffusion. bioRxiv. 2025 May 6:2025-05.

[14]. Broster JH, Popovic B, Kondinskaia D, Deane CM, Imrie F. Teaching Diffusion Models Physics: Reinforcement Learning for Physically Valid Diffusion-Based Docking. bioRxiv. 2026 Mar 27:2026-03.

[15]. Watson JL, Juergens D, Bennett NR, Trippe BL, Yim J, Eisenach HE, Ahern W, Borst AJ, Ragotte RJ, Milles LF, Wicky BI. De novo design of protein structure and function with RFdiffusion. Nature. 2023 Aug 31;620(7976):1089-100.

[16]. Baek M, DiMaio F, Anishchenko I, Dauparas J, Ovchinnikov S, Lee GR, Wang J, Cong Q, Kinch LN, Schaeffer RD, Millán C. Accurate prediction of protein structures and interactions using a three-track neural network. Science. 2021 Aug 20;373(6557):871-6.

[17*]. Masters MR, Mahmoud AH, Lill MA. Investigating whether deep learning models for co-folding learn the physics of protein-ligand interactions. Nature Communications. 2025 Oct 6;16(1):8854.

Through adversarial and perturbation tests, this study asks whether deep-learning co-folding models capture protein-ligand physics and concludes that current models largely interpolate within their training distribution rather than learning transferable physical interactions. It suggests caveats in treating such models as reliable physical oracles for design.

[18]. Feldman J, Brogi M, Skolnick J. Adversarial sequence mutations in AlphaFold and ESMFold reveal nonphysical structural invariance, confidence failures, and concerns for protein design. Computational and Structural Biotechnology Journal. 2026 Jun 29;35(1):0142.

[19]. Škrinjar P, Eberhardt J, Durairaj J, Schwede T. Have protein-ligand co-folding methods moved beyond memorisation?. BioRxiv. 2025 Feb 7:2025-02.

[20]. Karras T, Aittala M, Kynkäänniemi T, Lehtinen J, Aila T, Laine S. Guiding a diffusion model with a bad version of itself. Advances in Neural Information Processing Systems. 2024 Dec 16;37:52996-3021.

[21]. Geffner T, Didi K, Zhang Z, Reidenbach D, Cao Z, Yim J, et al.: Proteina: scaling flow-based protein structure generative models. In Proc 13th ICLR 2025.

[22]. Musaelian A, Batzner S, Johansson A, Sun L, Owen CJ, Kornbluth M, Kozinsky B. Learning local equivariant representations for large-scale atomistic dynamics. Nature Communications. 2023 Feb 3;14(1):579.

[23**]. Chen Y, Lu T, Zhao C, Wayment-Steele HK, Huang PS: SLAE: strictly local all-atom environment for protein representation. In Proc 43rd ICML 2026.

Introduces a strictly local all-atom environment representation learned with an asymmetric encoder-decoder in which energy prediction supervises a localised latent space, yielding physically meaningful, smoothly interpolable embeddings that transfer to downstream tasks such as NMR chemical-shift and function prediction.

[24]. Jacobs R, Morgan D, Attarian S, Meng J, Shen C, Wu Z, Xie CY, Yang JH, Artrith N, Blaiszik B, Ceder G. A practical guide to machine learning interatomic potentials–Status and future. Current Opinion in Solid State and Materials Science. 2025 Mar 1;35:101214.

[25]. Xia S, Zhang D, Shang X, Xu J. QuantaMind MD enables protein modeling with ab initio accuracy. bioRxiv. 2025 Sep 4:2025-09.

[26]. Li Z, Walsh A. Platonic representation of foundation machine learning interatomic potentials. Nature Machine Intelligence. 2026 May 7:1-1.

[27]. Edamadaka S, Yang S, Li J, Gómez-Bombarelli R. Universally converging representations of matter across scientific foundation models. arXiv preprint arXiv:2512.03750. 2025 Dec 3.

[28]. Passaro S, Corso G, Wohlwend J, Reveiz M, Thaler S, Somnath VR, Getz N, Portnoi T, Roy J, Stark H, Kwabi-Addo D. Boltz-2: Towards accurate and efficient binding affinity prediction. BioRxiv. 2025 Jun 18.

[29]. Russ WP, Figliuzzi M, Stocker C, Barrat-Charlaix P, Socolich M, Kast P, Hilvert D, Monasson R, Cocco S, Weigt M, Ranganathan R. An evolution-based model for designing chorismate mutase enzymes. Science. 2020 Jul 24;369(6502):440-5.

[30]. Li AJ, Lu M, Desta I, Sundar V, Grigoryan G, Keating AE. Neural network‐derived Potts models for structure‐based protein design using backbone atomic coordinates and tertiary motifs. Protein Science. 2023 Feb;32(2):e4554.

[31]. Shuai RW, Lu T, Bhatti S, Kouba P, Huang PS. Ensemble-conditioned protein sequence design with Caliby. bioRxiv. 2025 Oct 2:2025-09.

[32]. Socolich M, Lockless SW, Russ WP, Lee H, Gardner KH, Ranganathan R. Evolutionary information for specifying a protein fold. Nature. 2005 Sep 22;437(7058):512-8.

[33]. Marks DS, Hopf TA, Sander C. Protein structure prediction from sequence variation. Nature biotechnology. 2012 Nov;30(11):1072-80.

[34]. Kamisetty H, Ovchinnikov S, Baker D. Assessing the utility of coevolution-based residue–residue contact predictions in a sequence-and structure-rich era. Proceedings of the National Academy of Sciences. 2013 Sep 24;110(39):15674-9.

[35]. Fry B, Slaw K, Polizzi NF. Zero-shot design of drug-binding proteins via neural iterative selection− expansion. Nature. 2026 Jun 24:1-3.

[36]. El Nesr G, Dürr SL, Mathews II, Wen Q, Zhao K, Sarangi R, Röthlisberger U, Sunden F, Huang PS. Zero-shot design of a de novo metalloenzyme. bioRxiv. 2026 Apr 24.

[37]. Dürr SL, Levy A, Rothlisberger U. Metal3D: a general deep learning framework for accurate metal ion location prediction in proteins. Nature Communications. 2023 May 11;14(1):2713.

[38]. Dauparas J, Lee GR, Pecoraro R, An L, Anishchenko I, Glasscock C, Baker D. Atomic context-conditioned protein sequence design using LigandMPNN. Nature Methods. 2025 Apr;22(4):717-23.

[39]. Chu AE, Kim J, Cheng L, El Nesr G, Xu M, Shuai RW, Huang PS. An all-atom protein generative model. Proceedings of the National Academy of Sciences. 2024 Jul 2;121(27):e2311500121.

[40**]. Lauko A, Pellock SJ, Sumida KH, Anishchenko I, Juergens D, Ahern W, Jeung J, Shida AF, Hunt A, Kalvet I, Norn C. Computational design of serine hydrolases. Science. 2025 Feb 13;388(6744):eadu2454.

Reports the de novo design of catalytically active serine hydrolases, combining deep-learning backbone generation with PLACER to build preorganised, DFT-informed active sites and oxyanion holes. It demonstrates that catalysis could be distilled as conditioning information in protein structure generative models, while also exposing unmodelled components such as the activated water of the second reaction step.

[41]. Anishchenko I, Kipnis Y, Kalvet I, Zhou G, Krishna R, Pellock SJ, Lauko A, Lee GR, An L, Dauparas J, DiMaio F. Modeling protein–small molecule conformational ensembles with PLACER. Proceedings of the National Academy of Sciences. 2025 Nov 11;122(45):e2427161122.

[42]. Cao H, Quinn Z, Pal A, Kimura S, Zhang J, Heng PA, Chatterjee P. AlloGen: Conformation-Selective Binder Generation with Differential State Scoring. arXiv preprint arXiv:2606.05474. 2026 Jun 3.

[43]. Fried SD, Boxer SG. Electric fields and enzyme catalysis. Annual review of biochemistry. 2017 Jun 20;86:387-415.

[44]. Eberhart ME, Alexandrova AN, Ajmera P, Bím D, Chaturvedi SS, Vargas S, Wilson TR. Methods for theoretical treatment of local fields in proteins and enzymes. Chemical Reviews. 2025 Feb 24;125(7):3772-813.

[45]. Jabeen H, Beer M, Spencer J, Van Der Kamp MW, Bunzel HA, Mulholland AJ. Electric Fields Are a Key Determinant of Carbapenemase Activity in Class A β-Lactamases. ACS catalysis. 2024 Apr 23;14(9):7166-72.

[46]. Shen M, Dayhoff GW, Shen J. Protein Electrostatic Properties are Fine-Tuned Through Evolution. Research Square. 2025 Apr 28:rs-3.

[47]. Olsson MH, Søndergaard CR, Rostkowski M, Jensen JH. PROPKA3: consistent treatment of internal and surface residues in empirical p K a predictions. Journal of chemical theory and computation. 2011 Feb 8;7(2):525-37.

[48]. Braun M, Tripp A, Chakatok M, Kaltenbrunner S, Fischer C, Stoll D, Bijelic A, Elaily W, Totaro MG, Moser M, Hoch SY. Computational enzyme design by catalytic motif scaffolding. Nature. 2026 Jan 1;649(8095):237-45.

[49]. Kortemme T. De novo protein design—From new structures to programmable functions. Cell. 2024 Feb 1;187(3):526-44.

[50**]. Broerman AJ, Pollmann C, Zhao Y, Lichtenstein MA, Jackson MD, Tessmer MH, Ryu WH, Ogishi M, Abedi MH, Sahtoe DD, Allen A. Design of facilitated dissociation enables timing of cytokine signalling. Nature. 2025 Nov 13;647(8089):528-35.

Designs protein complexes in which dissociation kinetics are an explicit objective, engineering strained intermediates whose resolution by shear rather than rotation accelerates dissociation with applications to cytokine signalling timing. A demonstration of designing kinetics rather than only considering stable, ground states. Though it required manual specification and intuition of the strained states.

[51]. Haddox HK, Rocklin GJ, Motta FC, Strickland D, Halabiya SF, Cordray C, Park H, Klavins E, Baker D, DiMaio F. Using experimental results of protein design to guide biomolecular energy-function development. PLOS Computational Biology. 2026 Apr 22;22(4):e1014215.

[52]. Frey NC, Hötzel I, Stanton SD, Kelly R, Alberstein RG, Makowski EK, Martinkus K, Berenberg D, Bevers III J, Bryson T, Chan P. Lab-in-the-loop therapeutic antibody design with deep learning. BioRxiv. 2025 Feb 24:2025-02.

[53]. Huang W, Adornato GM, Horst M, Alturaifi TM, Hou K, Liu P, DeGrado WF, Yang Y. De Novo Design, Directed Evolution and Computational Study of Heme-Binding Helical Bundle Protein Catalysts for Biocatalytic Enantioselective Ge–H Insertion. Journal of the American Chemical Society. 2025 Oct 22;147(44):40869-78.

[54]. Martell CM, Gebis KK, Van HM, Gutierrez YM, Jung MD, Savas JN, Rocklin GJ. Global analysis of aggregation determinants in small protein domains. bioRxiv. 2025 Nov 12:2025-11.

[55*]. Ferrari ÁJ, Dixit SM, Thibeault J, Garcia M, Houliston S, Ludwig RW, Notin P, Phoumyvong CM, Martell CM, Jung MD, Tsuboyama K. Large-scale discovery, analysis and design of protein energy landscapes. Nature. 2026 May 13:1-1.

Measures residue-level energy profiles of conformational fluctuations for thousands of protein domains in parallel across two pH conditions, distinguishing cooperative folding from partially open intermediates via hydrogen-deuterium exchange. It can generate experimental energy-landscape data at a scale that could guide generative models to have conformational landscapes as an explicit design objective.

[56*]. Qian J, Milles LF, Wicky BI, Ragotte RJ, Motmaen A, Borst AJ, Skotheim R, Ols S, Coventry B, Li X, Kibler RD. Accelerating protein design by scaling experimental characterization. bioRxiv. 2025 Aug 6:2025-08.

A semi-automated, high-throughput pipeline that greatly increases the number of designed proteins that can be produced and experimentally characterized. It describes experimental infrastructure enabling rapid design-build-test-learn cycles at relatively low volume protein production scales.

[57]. Müntener T, Abramson D, Stern E, Hertel I, Jankevicius G, Mas G, Folkers GE, Wicky BI, Hiller S. Large-scale exploration of protein space by automated NMR. bioRxiv. 2026 Feb 16:2026-02.

[58]. Passow DR, Gupta A, Thompson S, Kundaje A, Fordyce PM. Amplicon/Protein Bead Display enables quantitative in vitro biochemistry at scale. bioRxiv. 2026:2026-05.

[59]. Cho Y, Tsuboyama K, Litberg TJ, Jung MD, Obisesan A, Wang Q, Phoumyvong CM, Thibeault J, Ovchinnikov S, Rocklin GJ. Accurate protein stability prediction for small domains using mega-scale experiments. bioRxiv. 2026 May 20:2026-05.

[60*]. Lewis S, Hempel T, Jiménez-Luna J, Gastegger M, Xie Y, Foong AY, Satorras VG, Abdin O, Veeling BS, Zaporozhets I, Chen Y. Scalable emulation of protein equilibrium ensembles with generative deep learning. Science. 2025 Jul 10;389(6761):eadv9817.

A generative model that emulates protein equilibrium ensembles trained across diverse protein structures, approximating Boltzmann-weighted conformational distributions orders of magnitude faster than molecular dynamics.

[61]. Abramson J, Adler J, Dunger J, Evans R, Green T, Pritzel A, Ronneberger O, Willmore L, Ballard AJ, Bambrick J, Bodenstein SW. Accurate structure prediction of biomolecular interactions with AlphaFold 3. Nature. 2024 Jun 13;630(8016):493-500.